\documentclass[prl,twocolumn,aps,letterpaper,preprintnumbers,superscriptaddress]{revtex4}
\usepackage{hyperref}
\usepackage{verbatim}
\usepackage{amsmath}
\usepackage{latexsym}
\usepackage{revsymb}
\usepackage{yfonts}
\usepackage{ifthen}
\usepackage{tcolorbox}
\usepackage{natbib}
\usepackage{amsfonts}
\usepackage{amsmath}
\usepackage{amssymb}
\usepackage{amsthm}
\usepackage{graphicx}
\usepackage{bm}
\usepackage{bbm}
\usepackage{epsfig,color,amssymb}
\usepackage{subfigure}
\usepackage{amsfonts}
\usepackage{amscd}
\usepackage{amsmath}
\usepackage{multirow}
\usepackage{chemarrow}
\usepackage{dcolumn}
\usepackage{bm}
\usepackage{graphicx}
\usepackage{enumerate}
\usepackage{epsfig}
\usepackage{subfigure}
\usepackage{xcolor}
\usepackage{multirow}
\usepackage{ulem}
\usepackage{braket}
\usepackage{comment}
\usepackage{enumitem}
\usepackage{amsthm}

\renewcommand{\emph}[1]{\textit{#1}}

\begin{document}

\title{Structured light analogy of squeezed state}

\author{Zhaoyang Wang}
\affiliation{Key Laboratory of Photonic Control Technology (Tsinghua University), Ministry of Education, Beijing 100084, China}
\affiliation{State Key Laboratory of Precision Measurement Technology and Instruments, Department of Precision Instrument, Tsinghua University, Beijing 100084, China}
\author{Ziyu Zhan}
\affiliation{Key Laboratory of Photonic Control Technology (Tsinghua University), Ministry of Education, Beijing 100084, China}
\affiliation{State Key Laboratory of Precision Measurement Technology and Instruments, Department of Precision Instrument, Tsinghua University, Beijing 100084, China}
\author{Anton N. Vetlugin}
\affiliation{Centre for Disruptive Photonic Technologies, SPMS, TPI, Nanyang Technological University, 637371 Singapore}
\author{Qiang Liu}
\affiliation{Key Laboratory of Photonic Control Technology (Tsinghua University), Ministry of Education, Beijing 100084, China}
\affiliation{State Key Laboratory of Precision Measurement Technology and Instruments, Department of Precision Instrument, Tsinghua University, Beijing 100084, China}
\author{Yijie Shen}\email{y.shen@soton.ac.uk}
\affiliation{Optoelectronics Research Centre, University of Southampton, Southampton SO17 1BJ, United Kingdom}
\author{Xing Fu}\email{fuxing@mail.tsinghua.edu.cn}
\affiliation{Key Laboratory of Photonic Control Technology (Tsinghua University), Ministry of Education, Beijing 100084, China}
\affiliation{State Key Laboratory of Precision Measurement Technology and Instruments, Department of Precision Instrument, Tsinghua University, Beijing 100084, China}
\date{\today}

\begin{abstract}
\noindent \textbf{Control of structured light is of great importance to explore fundamental physical effects and extend practical scientific applications, which has been advanced by accepting methods of quantum optics - many classical analogies of exotic quantum states were designed using structured modes. However, the prevailing quantum-like structured modes are limited by discrete states where the mode index is analog to the photon number state. Yet, beyond discrete states, there is a broad range of quantum states to be explored in the field of structured light -- continuous-variable (CV) states. As a typical example of CV states, squeezed state plays a prominent role in high-sensitivity interferometry and gravitational wave detection. In this work, we bring together two seemingly disparate branches of physics, namely, classical structured light and quantum squeezed state. We propose the structured light analogy of squeezed state (SLASS), which can break the spatial limit following the process of surpassing the standard quantum limit (SQL) with quantum squeezed states. This work paves the way for adopting methods from CV quantum states into structured light, opening new research directions of CV entanglement, teleportation, classical and quantum informatics of structured light in the future.}
\end{abstract}

\maketitle

\noindent Structured light has a great importance for developing both quantum and classical optical technologies, bridging the gap between quantum and classical optics and opening a platform for exchange of concepts and methodologies~\cite{forbes2021structured}. Many quantum effects could be reproduced by designing spatial modes of structured light~\cite{schott1927wave,cirac2012goals,qian2015shifting,shen2021creation}. For instance, the vortex beams carrying orbital angular momentum were exploited to simulate Landau levels and Laughlin states that unveil fermionic topological order in bosonic matters~\cite{schine2016synthetic,clark2020observation,corman2020light}. By manipulating a complex multi-vortex light field, the Berezinskii–Kosterlitz–Thouless phase transition was observed, opening the research of analogous thermodynamics in photonic light fluid~\cite{situ2020dynamics}. The vector beam with spatially unseparable polarization can resemble the quantum entangled Bell states, paving the way for exploitation of quantum measures for characterization of structured light ~\cite{kagalwala2013bell,forbes2019classically} and enabling quantum-inspired applications such as local teleportation~\cite{guzman2016demonstration},  turbulence-resilient communication and encryption~\cite{ndagano2017characterizing,zhu2021compensation}. Recently, vector vortex beams with multiple controlled unseparable degrees of freedom were designed for higher-dimensional entangled states~\cite{shen2021creation,shen2020structured,shen2021rays}, promising advanced applications from ultraprecise metrology to secure communications~\cite{shen2022nonseparable,wan2022divergence}.

However, the prevailing methodology mostly focused on the classical-quantum parallelism of structured light and discrete-variable (DV) quantum states, i.e. discrete eigenmodes of structured light (spatial mode indices, polarization) and discrete quantum number (energy level, photon spin). Further exploration is required to develop classical-qauntum analogy between structured light and continuous-variable (CV) quantum states, which are the states of variables with continuous spectrum such as position, momentum and amplitude of electromagnetic field. The best-known CV state is the squeezed state, where the uncertainty of one of the variables is suppressed compared to the coherent state~\cite{walls1983squeezed}. This allows to surpass the standard quantum limit (SQL) which is critical for high-sensitive metrology.

In this paper, we propose a generalized mathematical model for structured light analogy of squeezed state (SLASS), accompanied by paraxial wave equation (PWE). Typical forms of squeezed states, including squeezed vacuum states and squeezed number states, are mimicked by a new family of spatial structured modes. Analogues to quantum tomography techniques, we also assemble the experimental setup for generating and detecting the SLASS. Moreover, inspired by the superiority of squeezed state in surpassing the SQL, we demonstrate the SLASS overcoming spatial diffraction limits with deep subwavelength and superoscillation structures. The classical-quantum analogy, presented in this paper, allows bringing the concepts and methodologies of CV quantum optics into classical regime and develop novel approaches for such applications as super-resolution imaging and microscopy with classical light. At the same time, a new family of structured light can be utilized in quantum technologies such as structured photon manipulation, ultra-sensitive measurement, and quantum kernel computing~\cite{PhysRevLett.109.190403}.

\bigskip
\noindent
\textbf{\large Results}

\noindent \textbf{Concepts.}
In quantum optics, eigenstates of two modes with indices $j=1, 2$ are the solution of the stationary Schr\"{o}dinger equation, $\hat{\mathcal{H}} \left| n,m \right\rangle = E_{n,m} \left| n,m \right\rangle$, with the two-mode Hamiltonian operator $\hat{\mathcal{H}}=\frac{1}{2} \sum\limits_{j=1,2} \left( \hat{p}_j^2 + \hat{q}_j^2 \right)$~\cite{gerry2005introductory}. Here $\hat{q}_j = \left( \hat{a}_j + \hat{a}^{\dagger}_j \right) /\sqrt{2}$ and $\hat{p}_j = \left( \hat{a}_j - \hat{a}^{\dagger}_j  \right) /\left(\sqrt{2} \text{i}\right)$ are canonical variables satisfying commutation relation $\left[ \hat{q}_j,\hat{p}_j \right] = \text{i}$, $\hat{a}^{\dagger}$ and $\hat{a}$ are the creation and annihilation operators, $E_{n,m}$ is the energy of the state $\left| n,m \right\rangle$ which is the two-mode number state with $n$ photons in mode $1$ and $m$ photons in mode $2$ ($n$ and $m$ are integers). The set of all states $\left| n,m \right\rangle$ form a full orthonormal basis, allowing to express any pure two-mode quantum state as a superposition of number states. For instance, the two-mode squeezed vacuum state (SVS) can be represented as an infinite sum, $\sum\limits_{n} c_n \left| n,n \right\rangle$, where $c_n$ is the amplitude (Supplementary Material). Alternatively, this states can be represented as a squeezing operator acting on the two-mode vacuum state: $\hat{S}(\tau) \left|0,0 \right\rangle$ where $\tau$ is the squeezing parameter defining the magnitude and orientation of squeezing (Supplementary Material). Similarly, the displaced squeezed vacuum state (DSVS), the squeezed number state (SNS) and displaced squeezed number state (DSNS) can be expressed as $\hat{D}(\alpha) \hat{S}(\tau) \left|0,0 \right\rangle$, $\hat{S}(\tau) \left|0,N \right\rangle$ and $\hat{D}(\alpha) \hat{S}(\tau) \left|0,N \right\rangle$, respectively, where $N$ is an integer, $\hat{D}(\alpha)$ is the displacement operator with parameter $\alpha$ defining the magnitude and direction of displacement (Methods).

In classical optics, the Maxwell’s equations under paraxial approximation are reduced to the transverse eigenmode equation $\tilde{\mathcal{H}}^{\text{SL}} u_{n,m}(\xi,\eta) = C_{n,m}  u_{n,m} (\xi,\eta)$, where $C_{n,m}$ is the eigenvalue and $u_{n,m}(\xi,\eta)$ is the transverse mode function~\cite{gbur2011mathematical,Stoler:81}. The analytic form of $u_{n,m}(\xi,\eta)$ in the Cartesian coordinate is Hermite-Gaussian (HG) mode. Here, $\tilde{\mathcal{H}}^{\text{SL}}=\frac{1}{2} \sum\limits_{j=x,y} \left( {\tilde{p}}_j^2 + {\tilde{q}}_j^2 \right)$, $j=x,y$ represents two orthogonal directions in two-dimensional space, ${\tilde{p}}_j = -\text{i}\frac{\partial}{\partial j}$, ${\tilde{q}}_j = j$ satisfying the Poisson bracket relation, $\lbrace \tilde{q}_j,\tilde{p}_j \rbrace =1$. By denoting $u_{n,m}(\xi,\eta)$ as $|u_{n,m} \rangle$, we get $\tilde{\mathcal{H}}^{\text{SL}} |u_{n,m} \rangle = C_{n,m}  |u_{n,m} \rangle$ which has the form of the stationary Schrödinger equation above. The ``number state of structured light" $|u_{n,m} \rangle$ constitutes an infinite-dimensional space similar to the Hilbert space of quantum number states. In addition, it is worth noting that during structured light propagation, longitudinal coordinate $z$ could also be seen as analogy of time variable $t$ of quantum states (details in Supplementary Material).

Based on the similar mathematical framework, we construct the structured light analogy of CV quantum states. We use decomposition of CV quantum state in the basis of the number states, $\left| n,m \right \rangle$, and reproduce this superposition with the transverse eigenmode $|u_{n,m} \rangle$, where the amplitude of the state $\left| n,m \right \rangle$ is used as the amplitude of the mode $|u_{n,m} \rangle$. Similar to CV quantum states discussed above, we reproduce analogies of the SVS, DSVS, SNS and DSNS states - SLASS, using structured light. Following the methodology of quantum optics, these states can be represented via action of the corresponding operators to the transverse eigenmode: $ \tilde{S}(\tau) |u_{0,0} \rangle$, $ \tilde{D}(\alpha)\tilde{S}(\tau) |u_{0,0} \rangle$, $ \tilde{S}(\tau) |u_{0,N} \rangle$ and $ \tilde{D}(\alpha)\tilde{S}(\tau) |u_{0,N} \rangle$. Here the operators $\tilde{S}(\tau)$ and $\tilde{D}(\alpha)$ have the same form as their quantum counterparts, but applied to the states $|u_{n,m} \rangle$, where these analogous operators in structured light regime satisfy the same relation as their quantum counterparts (see Methods).

\begin{figure*}[t!]
	\centering
	\includegraphics[width=\linewidth]{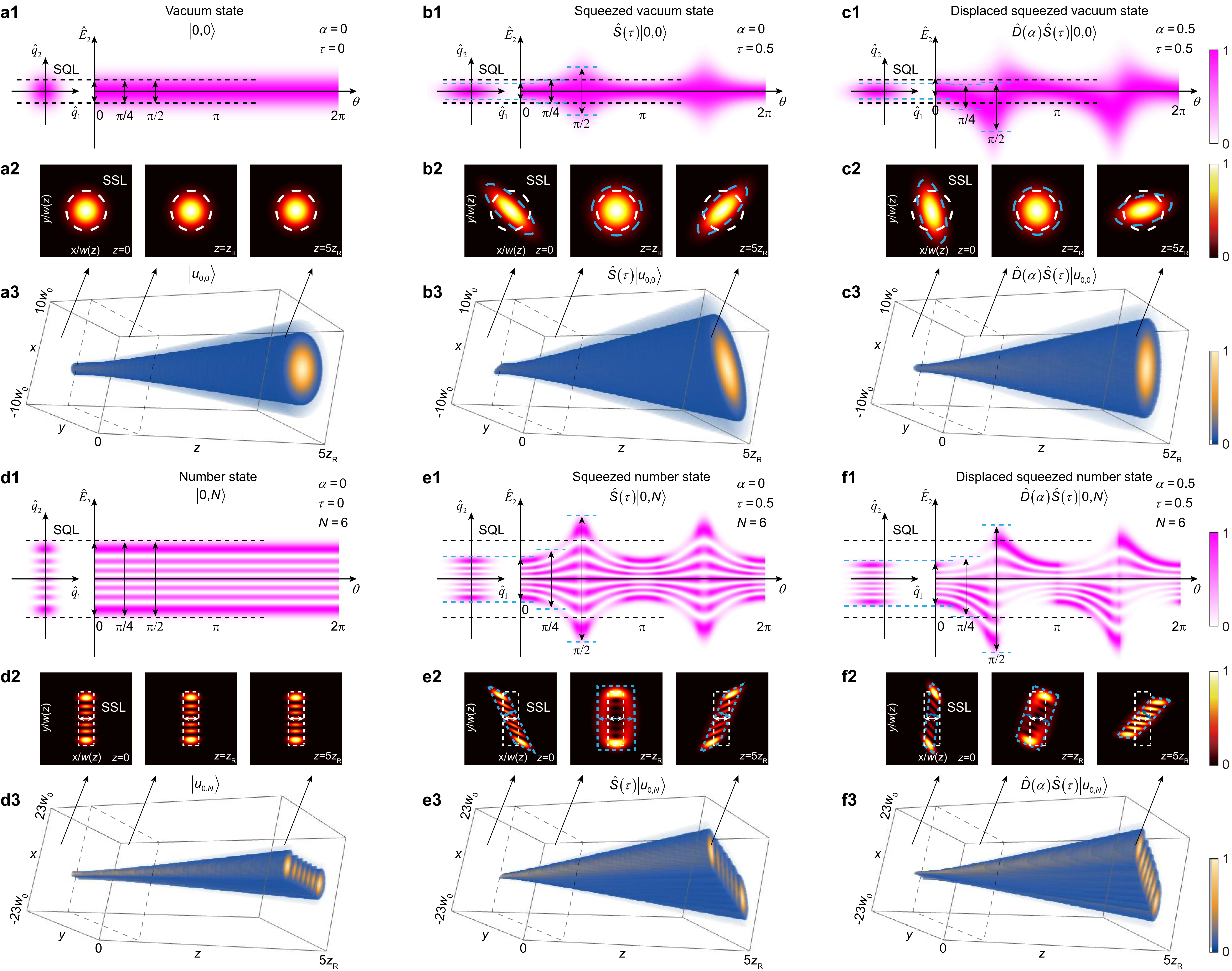}
	\caption{From quantum to structured light CV states. The subplots in the first and fourth rows are wave function for \textbf{a1} the vacuum state, \textbf{b1} SVS, \textbf{c1} DSVS, \textbf{d1} number state, \textbf{e1} SNS, \textbf{f1} DSNS, respectively. The gray dotted line marks the uncertainty for vacuum states corresponding to SQL, the blue dotted line marks the uncertainty for squeezed state corresponding to quantum noise, $\theta$ is in time dimension. The subplots in the second and fifth rows are transverse modes of SLASS (simulation) with different $z$, where $(x/w(z),y/w(z))$ ranges from $-5$ to $5$, $w(z)$ is the beam waist dependent on $z$. White dotted circles mark the waists of Gaussian beams and HG beams corresponding to SSL, blue dotted circles mark the waists of SLASS corresponding to spatial limit. The transverse modes with $z=5 z_{\text{R}}$ are approximated to the case of $z \to \infty$. The subplots in the third and sixth rows are the propagation of SLASS, where $z$ ranges from $0$ to $5z_{\text{R}}$, $(x,y)$ ranges from $-10w_0$ to $10w_0$ and $-23w_0$ to $23w_0$, respectively, $w_0$ is the beam waist with, $z_{\text{R}}$ is the Rayleigh range. SQL: standard quantum limit; SSL: standard spatial limit. Colormap: darkness to brightness means 0 to 1.}
	\label{theo}
\end{figure*}

\noindent \textbf{Structured light analogy of squeezed state.} 
Quantum two-mode vacuum state $\left|0,0 \right\rangle$ posses an isotropic uncertainty in the quadrature amplitudes, as shown in Fig.~\ref{theo} \textbf{a1}. Here we use the wave function to characterize the distribution of quadrature amplitudes of quantum state for a certain mode~\cite{schleich2011quantum}: the left subplot shows the wave function in phase space, and the right subplot shows the wave function in time evolution (varying $\theta$) on an arbitrary quadrature. The uncertainty of number state $\left|0,N \right\rangle$ is shown in Fig.~\ref{theo} \textbf{d1}. The wave function of SVS and SNS are shown in Fig.~\ref{theo} \textbf{b1} and \textbf{e1}, where the uncertainty is squeezed in the vertical direction but enlarged in the horizontal direction (left subplots). The uncertainty in the vertical direction also oscillates with varying $\theta$ at a period of $\pi$ (right subplots). The wave functions of DSVS and DSNS are overall displaced by the displaced operator, shown in Fig.~\ref{theo} \textbf{c1} and \textbf{f1}.

To develop the analogy of the CV quantum states in the structured light regime, we map the two-mode number state $\left| n,m \right \rangle$ to the HG$_{n,m}$ mode ($|u_{n,m} \rangle$). The two-mode quantum vacuum state is, thus, mapped to the fundamental HG$_{n,m}$ mode with $(n, m) = (0, 0)$, shown in Fig.~\ref{theo} \textbf{a2}. The transverse profiles of SLASS at various planes are shown in Fig.~\ref{theo} \textbf{b2}, \textbf{c2}, \textbf{e2} and \textbf{f2}, respectively. The beam waist of HG mode (marked with white dotted circles) can be seen as the counterpart for the SQL, and we name it as SSL. Similarly, the beam waist of SLASS can be seen as the counterpart for the quantum limit of quantum squeezed states, and we name it as the spatial limit. The spatial limits of SLASS are ellipses (marked with blue dotted circles), exhibiting anisotropic in different quadrature directions, analogous to the behavior of squeezed state in surpassing SQL.

Moreover, the longitudinal evolution of SLASS is illustrated in Fig.~\ref{theo} \textbf{a3}-\textbf{c3} and \textbf{d3}-\textbf{f3}, where the squeezed quadrature direction of structured light varies in propagation direction $z$. This behavior qualitatively reproduces that of the quantum squeezed state, where the direction of squeezing varies (oscillates) in time. Due to the diffraction of light, though, the distribution becomes broader for structured light as it propagates, which is not the case for squeezing of quantum light. 

\begin{figure}
	\centering
	\includegraphics[width=\linewidth]{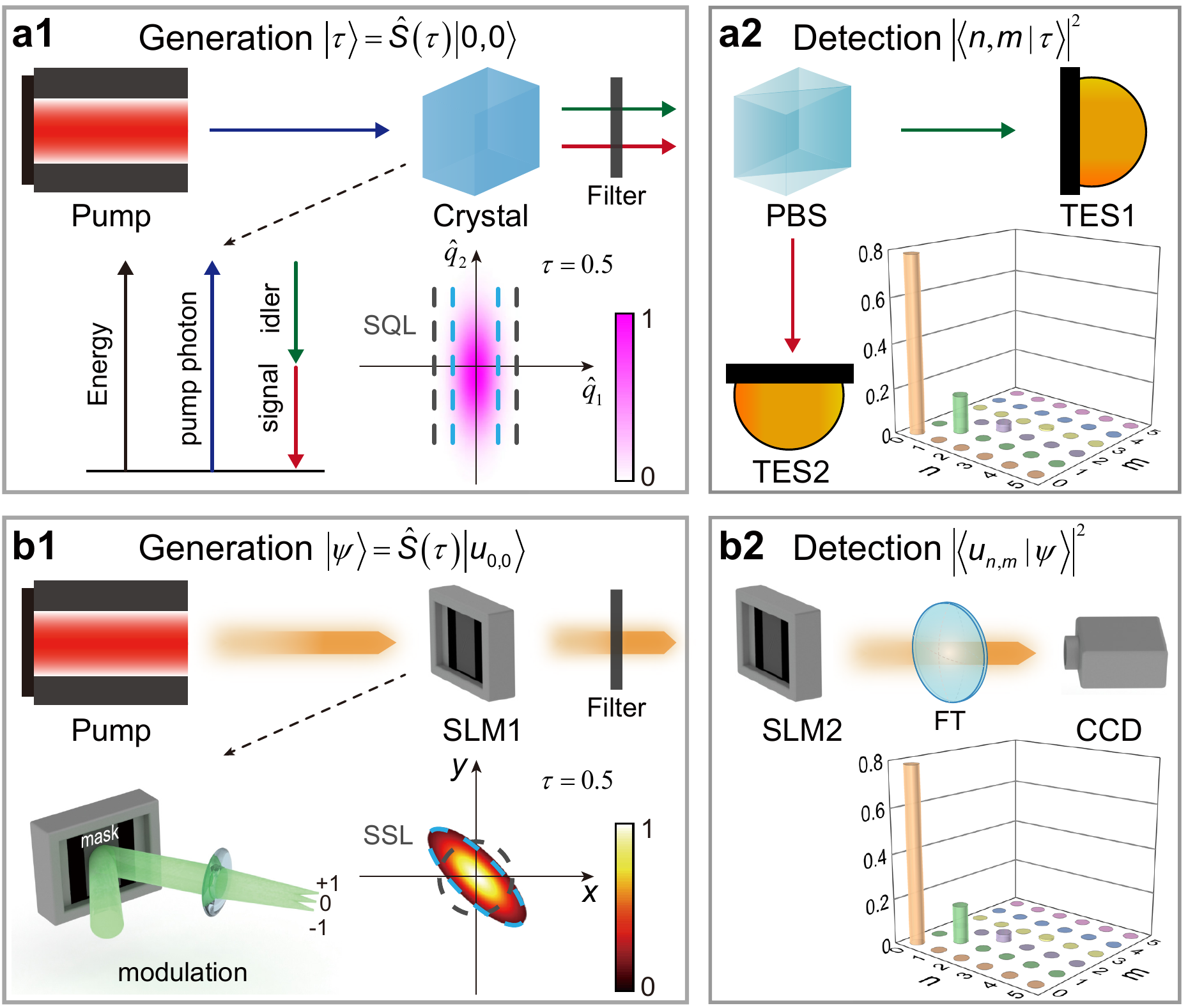}
	\caption{The schematic diagram of quantum-classical analogy between quantum squeezed state and SLASS. The generation (\textbf{a1}) and detection (\textbf{a2}) for SVS. The generation (\textbf{b1}) and detection (\textbf{b2}) for SLASS. PBS: polarizing beam splitter; TES: transition edge sensors; SLM: liquid-crystal spatial light modulator; FT: Fourier transform; CCD, charge coupled device camera.}
	\label{setup}
\end{figure}

\noindent \textbf{Quantum-classical analogy.}
In addition to the analogous mathematical descriptions, the SLASS can also exploit experimental methods of CV quantum optics. In this section, we show how the generation and measurement methods of quantum squeezed states can be replicated for SLASS (Fig.~{\ref{setup}}).

The two-mode squeezed vacuum state is routinely generated in the spontaneous parametric down-conversion process, where the non-linear crystal is pumped with a laser which is then filtered out to get access to the modes of interest, as shown in Fig.~{\ref{setup}} \textbf{a1} (here, the two modes differ in polarization). The SLASS is generated by sending a laser to the liquid crystal spatial light modulator (SLM1) with the following filtering, as shown in Fig.~{\ref{setup}} \textbf{b1}. The quantum-classical analogies for the generation setup are: SLM1 plays an analogous role as the crystal; there is a filtering process in both methods; the parametric down conversion process, shown in the bottom left of Fig.~{\ref{setup}} \textbf{a1}, is analogous to the light modulator process, shown in the bottom left of the Fig.~{\ref{setup}} \textbf{b1}; the wave function distribution of SVS with $\tau = 0.5$, shown in the bottom right of Fig.~{\ref{setup}} \textbf{a1}, corresponds to the SLASS, shown in the bottom right of the Fig.~{\ref{setup}} \textbf{b1}.

\begin{figure}
	\centering
	\includegraphics[width=\linewidth]{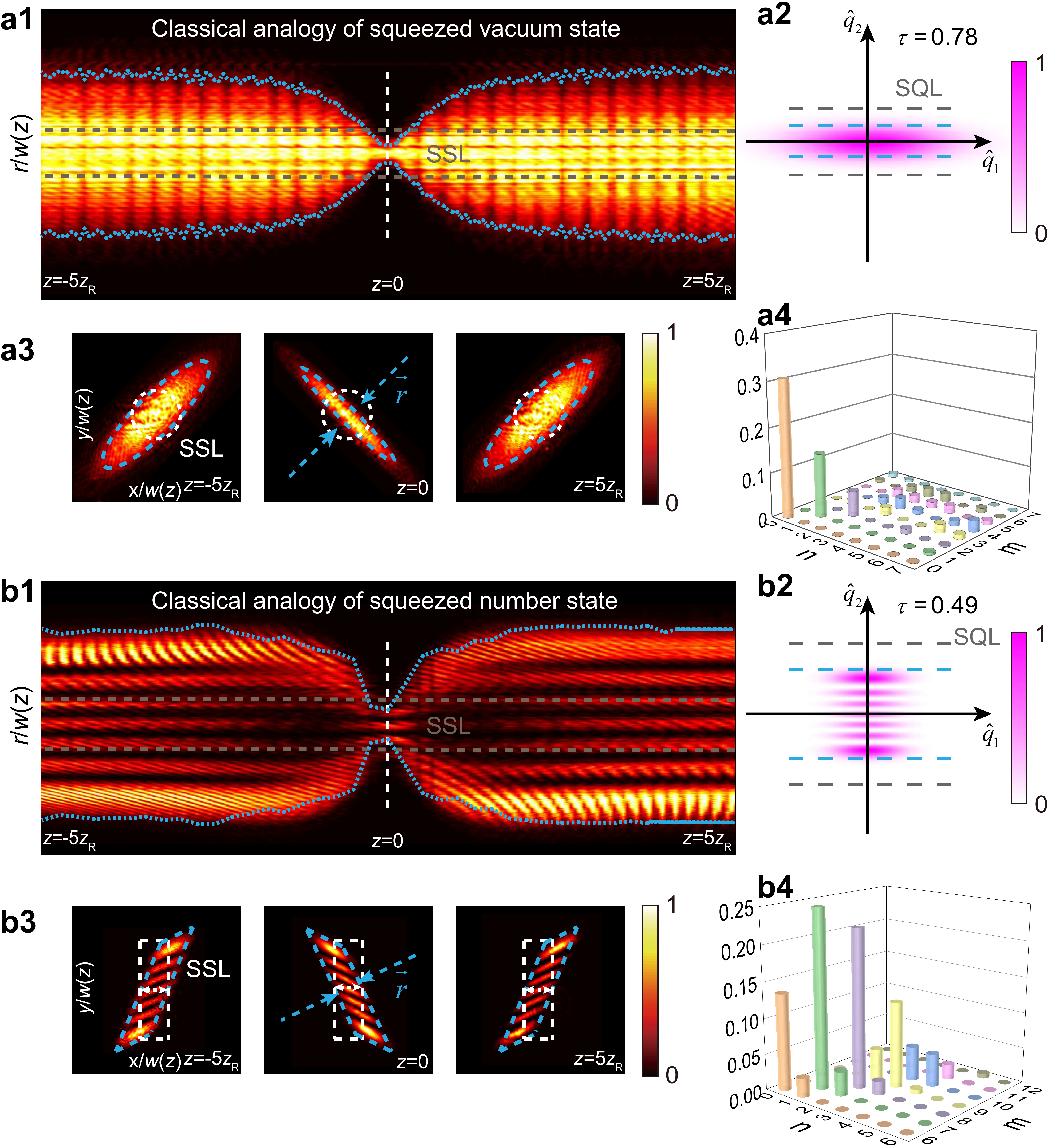}
	\caption{Surpassing SSL in experiment. The top and bottom sections are the cases of SVS and SNS of structured light, respectively. Subplots \textbf{a1} and \textbf{b1} are the intensity distributions at $\zeta-z$ plane, where $z$ ranges from $-5 z_{\text{R}}$ to $5 z_{\text{R}}$, $\zeta$ is the normalized radial position along the squeezed quadrature direction with $z=0$ (marked with blue arrows in subplots \textbf{a3} and \textbf{b3}). Subplots \textbf{a2}, \textbf{b2} are the corresponding wave function, with $\tau = 0.78$ and $\tau = 0.49$, respectively. Subplots \textbf{a3}, \textbf{b3} are the spatial modes with $z=-5 z_{\text{R}}$, 0, and $5 z_{\text{R}}$, respectively. Subplots \textbf{a4}, \textbf{b4} are the detected modal spectrums. The blue dotted lines (in \textbf{a1}, \textbf{a3}, \textbf{b1}, \textbf{b3}) marked the beam waist of SLASS. The gray dotted lines (in \textbf{a1}, \textbf{b1}) and white dotted lines (in \textbf{a3}, \textbf{b3}) marked the beam waists of the Gaussian beam and HG beam. Colormap: darkness to brightness means 0 to 1.}
	\label{exp3}
\end{figure}

The non-classical nature of the two-mode squeezed state is revealed through the photon-number correlation measurements~\cite{PhysRevLett.116.143601}, as shown in Fig.~{\ref{setup}} \textbf{a2}. Two modes are separated on a polarizing beam splitter (PBS) and each output of the PBS is detected by a photon-number resolving detector like the transition edge sensors (TES) to detect the state density $\left| \langle n,m|\tau \rangle \right| ^2$, where $|\tau \rangle = \hat{S}(\tau) |0,0 \rangle$ represents SVS. The detection process of SLASS, shown in Fig.~{\ref{setup}} \textbf{b2}, uses the second liquid-crystal spatial light modulator (SLM2) and CCD to detect the modal spectrum $\left| \langle u_{n,m}|\psi \rangle \right|^2$, where $|\psi \rangle$ represents SLASS. The SLM2 and CCD play the analogous roles as the PBS and TES, respectively. The simulation of state density and modal spectrum, with $\tau = 0.5$, is shown in the bottom right of the Fig.~{\ref{setup}} \textbf{a2} and \textbf{b2}, where they share the same distribution.

While generation and detection of the quantum SVS, in the experimental arrangement described above, is a routine procedure, there are, to the best of our knowledge, no such experimental demonstrations for the quantum SNS and DSNS as the setup would require substantial complications. In contrast, the generation and measurement of structured light counterparts of any CV states, including SNS and DSNS, can be implemented by simply changing the masks loaded on SLMs (see details about masks in \textbf{Methods}).

\begin{figure*}[t!]
	\centering
	\includegraphics[width=\linewidth]{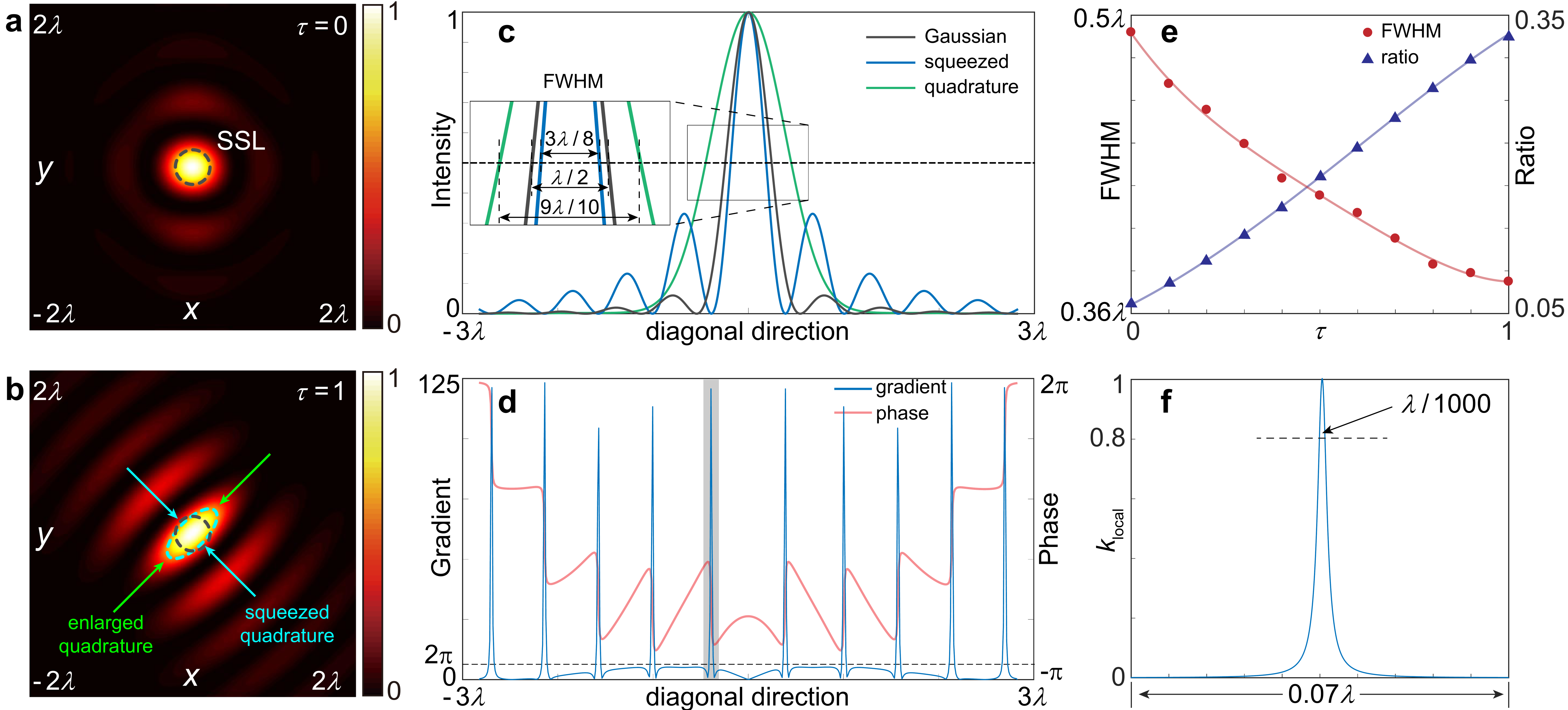}
	\caption{The simulation of tightly focused SLASS. Subplots \textbf{a} and \textbf{b} show the tightly focused modes with $\tau = 0$ and $\tau = 1$, respectively, where ($x$, $y$) ranges from $-2\lambda$ to $2\lambda$, blue and green arrows marked the two quadrature directions. Subplot \textbf{c} shows the intensity at the diagonal directions, where the black curve corresponds to the Gaussian beam with the FWHM $\lambda/2$, the blue and green curves corresponds to the intensity distributions at the squeezed and quadrature directions with the FWHM $3\lambda/8$ and $9\lambda/10$, respectively. Subplot \textbf{d} shows the phase and local wavenumber of structured light at the squeezed quadrature direction, corresponding to red and blue curves, respectively. Subplot \textbf{e} shows the FWHM and the relative intensity of the adjacent sideband to the central spot, with $\tau$ increasing from $0$ to $1$. Subplot \textbf{f} shows the zooming super-oscillatory structure, where the full width at $80\%$ maximum of local wavenumber is about $1/1000 \lambda$. Colormap: darkness to brightness means 0 to 1.}
	\label{fig5}
\end{figure*}

\noindent \textbf{Surpassing standard spatial limit.} 
Quantum squeezed states are most known for their ability to surpass the SQL. Remarkably, the structured light analogy of squeezed states, developed here, can be similarly used to surpass the SSL. In Fig.~\ref{exp3}, we demonstrate experimentally this application. Generally, the beam waists of structured light depend on propagation distance $z$. For convenience, we can also use the normalized radial position $\zeta = r/w(z)$, where $r$ represents the radial position along the squeezing direction at $z=0$, marked with blue arrow in Fig.~\ref{exp3} \textbf{a3} and \textbf{b3} for two distinct cases, respectively. The intensity of SLASS along the squeezed quadrature direction at $z=0$ during propagation could be clearly exhibited at a $\zeta-z$ plane, shown in Fig.~\ref{exp3} \textbf{a1} and \textbf{b1}. Fig.~\ref{exp3} \textbf{a2} and \textbf{b2} show the corresponding wave functions. Subplots \textbf{a3} and \textbf{b3} show the spatial modes at different $z$, presenting the varying squeezing direction of SLASS during propagation. Subplots \textbf{a4} and \textbf{b4} show the detected modal spectrum.

The spatial limit of structured light analogy of SVS is an ellipse, where the minor axis is the squeezing direction. The spatial limit at the squeezing direction is smaller than SSL. The squeezing direction varies in propagation, analogous to the oscillation of squeezing direction of quantum light. Similarly, one of the quadratures of the structured light analogy of SNS experiences squeezing while uncertainty of the other quadrature is enlarged. The squeezing direction also oscillates in propagation, qualitatively reproducing that of squeezed light.

\bigskip
\noindent
\textbf{\large Discussion.}
SLASS has potential applications in surpassing the SSL and providing a tunable source of super-oscillatory fields. We take the structured light analogy of SVS as the example to demonstrate this, as shown in Fig.~{\ref{fig5}}. Subplots \textbf{a}, \textbf{b} show the simulated tightly focused patterns of SLASS with $\tau=0$, $1$, respectively, $NA = 1.71$ selected for the simulation. Clearly, SLASS surpasses the SSL in the diagonal direction at the expense of the stretching in the anti-diagonal direction. The diagonal section of the intensity map of SLASS is shown in Fig.~{\ref{fig5}} \textbf{c}, where the full width at half maximum (FWHM) at the squeezing direction equals $3\lambda /8$, which is smaller than the FWHM of the tightly focused Gaussian beam, $\lambda /2$.  The FWHM of tightly focused SLASS at the anti-diagonal direction is $9\lambda /10$. Moreover, the tightly focused SLASS creates local wavenumbers exceeding the free-space wavenumber $2\pi/\lambda$, which leads to a super-oscillatory structure~\cite{yuan2019detecting}, as shown in Fig.~{\ref{fig5}} \textbf{d}. Here, the local wavenumber $k_{\text{local}}$, defined as the gradient of the phase $P(x,y)$, $k_{\text{local}} = |\nabla P(x,y) |$~\cite{yuan2019detecting}, is shown as the function of the position along the diagonal direction. Note that the dimension of the local wavenumber in Fig.~{\ref{fig5}} \textbf{d} is $1/\lambda$, and the super-oscillatory regions appear for $k_{\text{local}} > 2\pi$ (marked with a black dotted line). The super-oscillatory regions appear with the period of about $0.6\lambda$ and coincide with the regions of local minima of the field intensity. In general, there is a trade-off between the width and intensity of the super-oscillatory regions. For tightly focused SLASS, this trade-off is shown in Fig.~{\ref{fig5}} \textbf{e}: with increasing squeezing parameter $\tau$, the FWHM of the super-oscillatory spot decreases (red points and line) while the ratio of its adjacent sideband intensity to central spot intensity increases (blue triangles and line). The structure of the super-oscillatory region is shown in Fig.~{\ref{fig5}} \textbf{f}, where the full width at $80\%$ level is about $1/1000 \lambda$. 

As we have shown, the structured light analogy of squeezed states provides a robust method of surpassing the spatial diffraction limit and generation of super-oscillatory regions where both of these tools can be used for ultra-precise measurement~\cite{yuan2019detecting}, super-resolution imaging~\cite{Kozawa:18}, microscopy, etc. Beyond applications in classical optics, the complex spatial modes of SLASS can provide a substantial advantage for applications in quantum optics~\cite{SciHighNOONStates,yuan2019detecting,PhysRevLett.127.263601,SciSpatiallystructuredphotons}. For instance, interference of structured photons shaped into SLASS modes on a beamsplitter~\cite{PhysRevLett.126.123601} or in free space~\cite{PhysRevA.104.013716} would allow generating entangled states both in quantum and classical-analogy senses, producing highly correlated states of multi-dimensional structured light. Besides, SLASS modes of structured photons have multiple degrees of freedom, such as squeezing parameter $\tau$, displaced parameter $\alpha$, and mode number $N$, which could be exploited in high-dimension CV teleportation and communication~\cite{guzman2016demonstration,SciSpatiallystructuredphotons,wang2015quantum,science1227193}. Although we demonstrated SLASS modes in free space, these modes can exist in the multi-mode optical fibers and waveguides, parity-time (PT) and anti-PT symmetric Hamiltonian system~\cite{zivari2022non,sciadvabm7454}. Besides, the developed framework enables exploration of other classical analogy states of structured light, such as the cat state, NOON state, and GHZ state~\cite{Kozawa:18}, paving the way for further exploration of multi-dimensional structured light and its advanced applications.

\bigskip
\noindent
\textbf{\large Methods}

\noindent \textbf{The quantum operators and their structured light counterparts.} 
The ideal two-mode SVS $|\tau \rangle$ is defined via the quantum squeezed operator~\cite{gerry2005introductory}:
\begin{align}
    {\hat S}\left( \tau  \right) = \exp{ \left( {{\tau ^*}\hat{a}_1\hat{a}_2 - \tau {{\hat{a}_1}^{\dagger} }{{\hat{a}_2}^{\dagger} }} \right) }
	,\label{ss2}
\end{align}
where $\tau$ is a complex parameter, $\hat{a}_j$ and  $\hat{a}_j^{\dagger}$ ($j=1,2$) are the quantum annihilation and creation operators as $ \hat{a}_1 \hat{a}_2 |n,m \rangle = \sqrt{nm} |n-1,m-1 \rangle$, $\hat{a}^{\dagger}_1 \hat{a}^{\dagger}_2 |n,m \rangle = \sqrt{(n+1)(m+1)} |n+1,m+1 \rangle$, respectively. SVS is defined as $|\tau \rangle = \hat{S}(\tau)  |0,0 \rangle$, which cloud be decomposed into a superstition of number states as~\cite{gerry2005introductory}:
\begin{align}
    |\tau \rangle = \frac{1}{{\cosh |\tau|}}\sum\limits_{K = 0}^\infty  {{\left( { - 1} \right)^K}\text{e}^{\text{i} K \theta }  { {\tanh^K |\tau|}}\left| {K,K} \right\rangle } ,\label{ss3} 
\end{align}
where $\theta = \arg \tau$. SVS exhibits strong correlations between the modes, as only paired states $\left| {K,K} \right\rangle$ exist in the state $|\tau \rangle$. This non-classical correlation between the two modes can be measured experimentally and be used for verification of the non-classical squeezed state, as shown in Fig.~{\ref{setup}} \textbf{a2}. The more general states, DSVS and DSNS, are defined via the quantum displaced operator~\cite{gerry2005introductory}:
\begin{align}
    {\hat D}\left( \alpha  \right) = \exp{ \left( {{\alpha}\hat{a}_1^{\dagger}\hat{a}_2 - \alpha^* {\hat{a}_1}{{\hat{a}_2}^{\dagger} }} \right) }
	,\label{d2}
\end{align}
where $\alpha$ is also a complex parameter. Both $\hat{S}(\tau)$ and $\hat{D}(\alpha)$ are defined via annihilation $\hat{a}_j$ and creation $\hat{a}_j^{\dagger}$ operators, which could be defined via the canonical variable operators $\hat{p}_j$ and $\hat{q}_j$ as~\cite{gerry2005introductory}:
\begin{align}
	\hat{a}_j = \frac { \hat{q}_j + \text{i}\hat{p}_j } {\sqrt{2}},\label{a}
\end{align}
\begin{align}
	\hat{a}_j^\dagger =\frac { \hat{q}_j - \text{i}\hat{p}_j } {\sqrt{2}},\label{a+}
\end{align}
where $\hat{p}_j$ and $\hat{q}_j$ correspond to observable quantities, related to the operators of the electric and magnetic fields as:
\begin{align}
	\hat{E}_{j} = \sqrt{ \frac { 2 \omega^2 } {V \varepsilon_0} } \hat{q}_j \sin \left( kz \right),\label{Ex}
\end{align}
\begin{align}
	\hat{B}_{j} = \left( \frac{ \mu_0 \varepsilon_0 }{k} \right) \sqrt{ \frac { 2 \omega^2 } {V \varepsilon} } \hat{p}_j \cos \left( kz \right),\label{By}
\end{align}
where $\hat{E}_{j}$ and $\hat{B}_{j}$ are the electric and magnetic fields operators of the $j$ mode, $V$ is the effective volume of the electric and magnetic fields, $\varepsilon_0$ is the vacuum Permittivity, $\mu_0$ is the vacuum permeability. Since we usually focus on the electric field of light, the first and fourth rows of Fig.~\ref{theo} only exhibit the wavefunctions of quantum states in the position bases $\left\{ \hat{q}_j \right\}$ and the varying $\hat{E}_{2}$ in time evolution. The electric operator $\hat{E}_{j}$ corresponds to the transverse intensity distribution of structured light, which are shown in the second and fifth rows of Fig.~\ref{theo}.

The structured light counterparts of annihilation and creation operators are defined as:
\begin{align}
	\tilde{a}_j = \frac { \tilde{q}_j + \text{i}\tilde{p}_j } {\sqrt{2}},\label{b}
\end{align}
\begin{align}
	\tilde{a}_j^\dagger =\frac { \tilde{q}_j - \text{i}\tilde{p}_j } {\sqrt{2}},\label{b+}
\end{align}
where ${\tilde{p}}_j = -\text{i}\frac{\partial}{\partial j}$, ${\tilde{q}}_j = j$ in Cartesian coordinate, here $j=x,y$ represents two orthogonal directions in two-dimensional space. The affection of $\tilde{a}_j$ and $\tilde{a}_j^\dagger$ on $\left| u_{n,m} \right\rangle$ is similar to that of the quantum annihilation and creation operators as $ \tilde{a}_1 \tilde{a}_2 |u_{n,m} \rangle = \sqrt{n m} |u_{n-1,m-1} \rangle$, $\tilde{a}^{\dagger}_1 \tilde{a}^{\dagger}_2 |u_{n,m} \rangle = \sqrt{(n+1)(m+1)} |u_{n+1,m+1} \rangle$.

The structured light counterparts of squeezed and displaced operators are defined as:
\begin{align}
    {\tilde S}\left( \tau  \right) = \exp{ \left( {{\tau ^*}\tilde{a}_1\tilde{a}_2 - \tau {{\tilde{a}_1}^{\dagger} }{{\tilde{a}_2}^{\dagger} }} \right) }
	,\label{cs2}
\end{align}
\begin{align}
    \tilde{ D}\left( \alpha  \right) = \exp{ \left( {{\alpha}\tilde{a}_1^{\dagger}\tilde{a}_2 - \alpha^* {\tilde{a}_1}{{\tilde{a}_2}^{\dagger} }} \right) }
	.\label{cd2}
\end{align}

The quantum operators and their structured light counterparts share the same mathematical forms, thus the SLASS could be decomposed into the same superstition state. For example, the structured light analogy of SVS, $\left|\psi \right \rangle = \tilde{S}(\tau) \left|u_{0,0} \right \rangle$, could be decomposed as:
\begin{align}
    |\psi \rangle = \frac{1}{{\cosh |\tau|}}\sum\limits_{K = 0}^\infty  {{\left( { - 1} \right)^K}\text{e}^{\text{i} K \theta }  { {\tanh^K |\tau|}}\left| u_{K,K} \right\rangle } 
    ,\label{cs3} 
\end{align}
where only transverse eigenmodes with same indices $\left| u_{K,K} \right\rangle$ exist in the structured light counterpart $|\psi \rangle$, exhibiting strong correlations between the eigenmodes as quantum squeezed states. This analogous correlation can be measured experimentally via the setup shown in Fig.~{\ref{setup}} \textbf{b2}.

\noindent \textbf{Generation and detection of squeezed state of structured light.} 
A liquid-crystal spatial light modulator (SLM1) was exploited to experimentally generate the SLASS~\cite{Arrizon:07}. Liquid-crystal SLM is a kind of opto-electrical device for phase modulation, which is regulated by the extraordinary refractive index of liquid crystal cells. This method requires masks, which are computer-generated holograms, essentially~\cite{Arrizon:07}. The masks could be expressed as~\cite{Arrizon:07}:
\begin{align}
	f_{\text{mask}}(x,y)=\text{e}^{\text{i} J_1^{-1}[CA(x,y)] \sin[P(x,y)+2\pi(u_x x + v_y y)]},
	\label{mask}
\end{align}
where $C=0.5819$ is a constant, $A(x,y)$ and $P(x,y)$ are the complex amplitude and phase of target lights, $(u_x,u_y)$ is the spatial frequency coordinate. The beam emitted by the pump would be expanded and collimated before SLM1. Then the collimated beam is modulated by the mask loaded on SLM1. There are several order components in the modulated beam, which would go through a 4$f$ system and be filtered in the Fourier plane to extract the +1th-order component, as shown in the bottom left insert of Fig.~{\ref{setup}} \textbf{b1}. The filtered beam is the SLASS $|\psi \rangle$. To measure the modal spectrum of SLASS, SLM2 is used and loaded the mask of $\langle u_{n,m} |$, where the wave function corresponding to $\left \langle u_{n,m} \right|$ is the conjugate of the corresponding wave function of $\left| u_{n,m} \right \rangle$. The modulated light goes through the Fourier transform by a lens (labelled "FT" as shown in the detection part of Fig.~{\ref{setup}} \textbf{b2}). The modal spectrum information $\left| \langle u_{n,m} | \psi \rangle \right|^2$ exists in the center of +1th-order component, the intensity of its central spot, which could be detected by CCD. The complete modal spectrum could be obtained just by changing the mask of $\langle u_{n,m} |$ with various indices and repeating this process.

$| u_{n,m} \rangle $ and $| \psi \rangle $ only describe the transverse pattern of structured light, which should be added a propagation factor to characterize a real beam, represented as $U_{n,m}(x,y,z)$ and $\Psi(x,y,z)$. Therefore, what we measured in experiment is $b_{n,m} = \iint U^{*}_{n,m}(x,y,z) \Psi(x,y,z) \mathrm{d} x \mathrm{d} y $. The value of $b_{n,m}$ is the same as the $ \left| \langle u_{n,m} | \psi \rangle \right|^2$ for any $z$, i.e. the modal spectrum of real beams is independent on $z$. In this work, we measured the modal spectrums of SLASS at $z=0$ plane. 

\noindent\textbf{Tightly focusing simulation.}
The limit of classical light is realized in tightly focused beams. To illustrate the property of surpassing the diffraction limit for SLASS, we numerically simulate the behavior of SLASS after being focused by a large numerical aperture lens. The focused electromagnetic field could be calculated via non-paraxial vector diffraction theory based on Debye approximation, which is expressed by angular spectrum representation as~\cite{novotny2012principles}:
\begin{equation}
\begin{split}
	\mathbf{E}(\rho,\varphi_2,z) = 
	& \frac{\text{i} k f \text{e}^{-\text{i} k f}}{2\pi}\int_{0}^{\vartheta_{max}}\int_{0}^{2\pi}\mathbf{\tilde{E}}_{\text{input}}(\vartheta,\varphi_1) \text{e}^{\text{i} k z \cos \vartheta} \\
	& \text{e}^{\text{i} k \rho \sin\vartheta \cos(\varphi_1-\varphi_2)} \sin \vartheta \mathrm{d}\vartheta \mathrm{d}\varphi_1
	\label{diff}
\end{split}
\end{equation}
where $f$ is the focal length, $k$ is the wavenumber, $\rho$ and $\varphi_2$ represent the radial and azimuthal coordinates in the focused plane, $\varphi_1$ is the azimuthal coordinate in the input plane, $z$ is the longitudinal distance between the focused plane and the input plane, $\vartheta$ is the incident angle where the maximum incident angle $ \vartheta_{max}$ is determined from the numerical aperture (NA) as $ NA=n_{1}\sin(\vartheta_{max})$, $n_{1}$ and $n_{2}$ are the refractive indices of media in the object and image space. In addition, $ \mathbf{E}_{\text{input}}(\vartheta,\varphi_1)$ is the input electric field, modulated by a large numerical aperture lens as~\cite{novotny2012principles}:
\begin{equation}
\begin{split}
& \mathbf{\tilde{E}}_{\text{input}}(\vartheta,\varphi_1)=  \\
& t^{s}(\vartheta) \left[ \mathbf{E_{\text{input}}} \cdot  \left(
\begin{aligned}
-    \sin(&\varphi_1) \\
    \cos(&\varphi_1) \\
    0
\end{aligned}
\right) \right]
\left(
\begin{aligned}
-    \sin(&\varphi_1) \\
    \cos(&\varphi_1) \\
    0
\end{aligned}
\right) \sqrt{\frac{n_{1}}{n_{2}}} \sqrt{\cos{\vartheta}} + \\
& t^{p}(\vartheta) \left[ \mathbf{E_{input}} \cdot  \left(
\begin{aligned}
    \cos(&\varphi_1) \\
    \sin(&\varphi_1) \\
    0
\end{aligned}
\right) \right]
\left(
\begin{aligned}
   \cos(\varphi_1)\cos(&\vartheta) \\
    \sin(\varphi_1)\cos(&\vartheta) \\
   -\sin(&\vartheta)
\end{aligned}
\right) \sqrt{\frac{n_{1}}{n_{2}}} \sqrt{\cos{\vartheta}}
\label{diff2}
\end{split}
\end{equation}
where $\mathbf{E_{\text{input}}}$ is the original input electric field, $t^{s}(\vartheta)$ and $ t^{p}(\vartheta)$ are the transmission coefficient of the interface at the focal lens plane for $s-$ and $p-$ polarized beam, respectively. Based on the Eq.~(\ref{diff}) and Eq.~(\ref{diff2}), it is possible to acquire the relatively accurate optical field of any position behind the focal lens.
\bibliographystyle{naturemag}

\bigskip
\noindent\textbf{\large Acknowledgments}\\
This work is funded by National Natural Science Foundation of China (61975087); Marie S.-Curie MULTIPLY Fellowship (GA713694); the Singapore Ministry of Education [No. MOE2016-T3-1-006 (S)].

\bigskip
\noindent\textbf{\large Competing interests}\\
The authors declare no competing interests.

\bibliography{sample}

\end{document}